\documentclass[conference]{IEEEtran}

\ifCLASSINFOpdf
\else
\fi
\usepackage{multicol}
\usepackage{tikz}
\usepackage{tkz-graph}
\usepackage{bbm}
\usepackage{booktabs}
\usepackage{algpseudocode}
\usepackage{algorithm}

\usepackage{color}
\usepackage[cmex10]{amsmath}
\usepackage{dblfloatfix}
\usepackage{amsfonts}
\usepackage{url}
\usepackage{pgfplots}
\usepackage{pgfplotstable}
\usepackage{hyperref}
\hypersetup{
    colorlinks=true,
    linkcolor=blue,
    filecolor=magenta,      
    urlcolor=cyan,
}
\begin{document}
\pagenumbering{gobble}

%
\title{\textbf{\Large An Open-Source Integration of Process Mining Features into the Camunda Workflow Engine: Data Extraction and Challenges\\[-1.5ex]}}

\author{
\IEEEauthorblockN{~\\[-0.4ex]\large Alessandro Berti\IEEEauthorrefmark{1}, Wil van der Aalst\IEEEauthorrefmark{1},
David Zang\IEEEauthorrefmark{2}, Magdalena Lang\IEEEauthorrefmark{2} \\[0.3ex]\normalsize}
\IEEEauthorblockA{\IEEEauthorrefmark{1}Process and Data Science Department, RWTH Aachen University\\
Process and Data Science department, Lehrstuhl fur Informatik 9 52074 Aachen, Germany\\
Emails: {a.berti@pads.rwth-aachen.de}, {wvdaalst@pads.rwth-aachen.de}}
\IEEEauthorblockA{\IEEEauthorrefmark{2}viadee Unternehmensberatung AG\\
Konrad-Adenauer-Ufer 7, 50668\\
Emails: {david.zang@viadee.de}, {magdalena.lang@viadee.de}}
}

\maketitle

\begin{abstract}
Process mining provides techniques to improve the performance and compliance of operational processes.
Although sometimes the term ``workflow mining'' is used, 
the application in the context of 
Workflow Management (WFM) and Business Process Management (BPM) systems is limited.
The main reason is that WFM/BPM systems control the process, leaving less room 
for flexibility and the corresponding deviations. 
However, as this paper shows, it is easy to extract event data from systems like Camunda, 
one of the leading open-source WFM/BPM systems.
Moreover, although the respective process engines control the process flow, process mining is still able to provide valuable insights, 
such as the analysis of the performance of the paths and the mining of the decision rules.
This demo paper presents a process mining connector to Camunda that extracts event logs and process models, allowing for the application of existing process mining tools. 
We also analyzed the added value of different process mining techniques in the context of Camunda.
We discuss a subset of process mining techniques that nicely complements the 
process intelligence capabilities of Camunda.
Through this demo paper, we hope to boost the use of process mining among Camunda users.
\end{abstract}




%
\IEEEpeerreviewmaketitle

\section{Introduction}

\begin{figure*}[!t]
\centering
\includegraphics[width=0.70\textwidth]{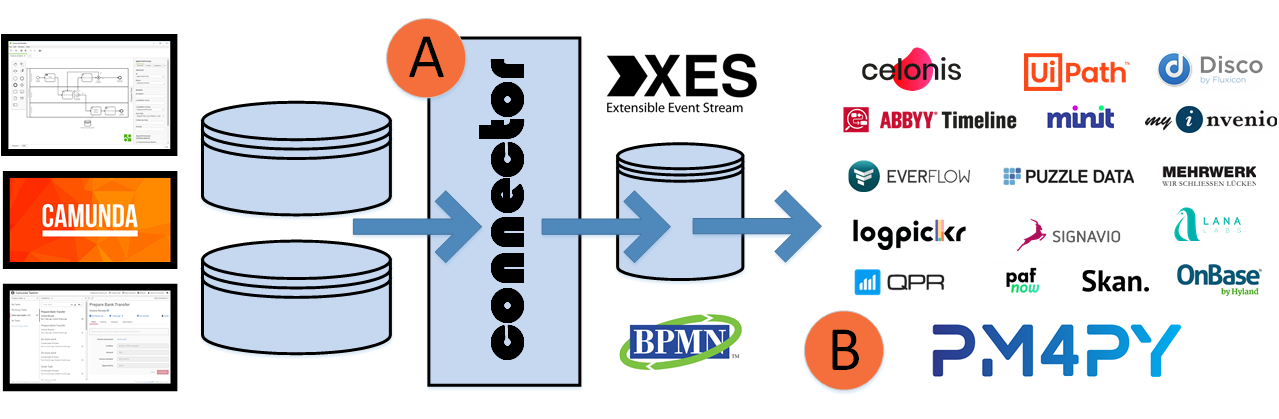}
\caption{Overview of the toolchain supporting process mining in the context of Camunda.
In this paper, we provide
(A) a connector to Camunda that is able to extract event logs and 
the BPMN diagrams modeling the processes.
(B) an overview on the most valuable process mining techniques complementing Camunda.
Although the connector is generic, we showcase the integration using PM4Py.}
\label{fig:introduction}
\end{figure*}

The vast majority of business processes (including enterprise resource planning, customer relationship management, document management) 
are nowadays supported by information systems. 
These systems manage (but not always regulate) the execution of a business process, and
record {\it event data} with fine detail about each step of the process. 
In this context, process mining \cite{van2011process} allows to improve operational processes by exploiting
the event data recorded by such systems. An {\it event log} can be extracted from an information system's database in order to apply the process mining algorithms.
An event log contains event data of multiple executions of the business process.
Process mining techniques include: {\it process discovery}, i.e., the automated discovery of a process model from event data;
{\it conformance checking}, i.e., the comparison between a process model and the event data; {\it model enhancement}, i.e.,
the enrichment of the model with additional perspectives (for example, execution guards \cite{de2013data}), prediction and
simulation algorithms. Open-source software supporting process mining includes ProM, APROMORE
and PM4Py.

Process mining techniques have been applied to Workflow Management (WFM) and Business Process Management (BPM) systems.
There are connectors to the YAWL WFM system \cite{rozinat2008workflow} and some
other WFM/BPM systems, including Signavio, Bizagi, and Bonita, that allow to to extract the event data and operate on the process models
contained in such systems.
This paper focuses on Camunda and is a result of a collaborative project between the RWTH Aachen University
and {\it viadee Unternehmensberatung AG}. 
Before, there was no open-source connector to extract logs useful for process mining purposes from Camunda, 
although Camunda is open-source and holds detailed event data.
Therefore, we developed and evaluated such a connector, and
{\it viadee} has integrated event log extraction techniques in its software stack.

Camunda is widely used, e.g., by Deutsche Telekom, Warner Music, Allianz, DB, Zalando and Generali.
Camunda uses the BPMN 2.0 notation for modeling.
Among the main selling points of Camunda are high throughput and 
collaboration and integration possibilities.
Camunda can be easily integrated with different information systems, 
business intelligence and big data systems such as
QlikView, Apache Spark and Kafka. 
This explains our goal to provide process mining for the large Camunda user base.

In this demo paper,
(A) we present our implementation of a process mining extractor for Camunda, that is able to extract
a set of event logs for the processes executed by Camunda, and
(B) we discuss the existing process mining techniques that complement the business intelligence capabilities of Camunda. Figure~\ref{fig:introduction} provides an overview
of the approach implemented in the paper. 
The extractor is publicly available. For demonstrative purposes, it is integrated 
with a graphical interface based on PM4Py that offers
basic process mining functions. Moreover, the techniques analyzed in this paper are available in open-source software. 

The remainder of this demo paper is organized as follows. Section~\ref{sec:extractingEventLogsFromCamunda} describes the basic structure of the Camunda database,
along with a methodology of extraction of event logs and process models from the Camunda database. 
Section~\ref{sec:processMiningCamunda} discusses the added value of process mining
techniques for Camunda users.
Section \ref{sec:setup} describes the set-up of the tool. Finally,
Section \ref{sec:conclusion} concludes the paper.

\section{Extracting Event Logs and Process Models From Camunda}
\label{sec:extractingEventLogsFromCamunda}

In this section, we will focus on how to extract logs containing the historical executions of the processes supported by Camunda,
and how to extract the process models of such processes.

\subsubsection{Extracting Event Logs}

The extraction is done directly at the database level.
The Camunda workflow engine supports different relational databases (e.g., PostgreSQL, Oracle, MySQL).
We will focus exclusively on the completed executions, and ignore ongoing executions.

The table containing the historical executions of the processes is the ACTI\_HI\_ACTINST table. The rows of this table are the events thrown by Camunda.
The table contains all the basic information that is
needed to extract event logs:
\begin{itemize}
\item The identifier of the process that is executed is stored in the {\it proc\_def\_key\_} column. This column contains as many different values as processes are executed via the Camunda process engine.
\item The identifier of the process execution (case ID) is stored inside the {\it proc\_inst\_id\_} column.
\item The name of the BPMN element that are executed via the Camunda process engine is stored inside the {\it act\_name\_} column.
\item The type of the BPMN element is stored inside the {\it act\_type\_} column.
\item The start and end timestamps are stored inside the {\it start\_time\_} and the {\it end\_time\_} columns, respectively.
\item The identifier of the resource that performs the event is stored inside the {\it assignee\_} column.
\end{itemize}
Basically, an event log is created for each distinct value of the {\it proc\_def\_key\_} column.
The resulting table for an individual process is enough
to analyze the control flow of the process and its bottlenecks.
Other attributes at the event level can be obtained by merging the {\it ACT\_HI\_ACTINST} table with the {\it ACT\_HI\_DETAIL} table. The latter contains a row for each distinct attribute that is associated with an event.
These attributes can be useful to investigate the process more thoroughly, also for predictive analyses.

\begin{table*}[!t]
\caption{Analysis of the pros and cons of the application of several process mining techniques in the context of the Camunda  engine.
Many observations here hold generally for any WFM/BPM system.}
\centering
\vspace{-2mm}
\begin{tabular}{|p{3cm}|p{6cm}|p{6cm}|}
\hline
{\bf Technique} & {\bf Pros} & {\bf Cons} \\
\hline
Process Discovery & It is possible to show the frequent paths in processes. Moreover, it becomes visible when people bypass the system. & The process model underlying the event data is already contained in Camunda and probably not surprising. \\
\hline
Conformance Checking & It is possible to measure the precision of the process model, in order to understand how much extra behavior is allowed. & It is expected that the event data already follows the model. Hence, some measures such as the calculation of fitness are not useful. \\
\hline
Decision Mining & It is possible to enrich the BPMN model with guards that describe and regulate the behavior of the process at the decision points. & Many execution guards are already inserted in the BPMN diagrams during the design phase. The discovered guards might be trivial or overfit the data. \\
\hline
Concept Drift Analysis & Process mining can be used to detect process changes. Possible reasons include changes of the process model or day-night shifts. & Many of the possible change points are known or deliberate. \\
\hline
Prediction of the Remaining Time & The technique provides an estimation of the remaining time for the process instances, in order to detect possible service level agreements violations. & The quality of the predictions performed 
by state-of-the-art approaches on real datasets must be improved. \\
\hline
Social Network Analysis & The collaboration between the resources can be analyzed from different angles (e.g., to see the effect on performance).
& Roles are often set and controlled by the system. \\
\hline
Model Enhancement & It is possible to identify the bottlenecks of the process, and the most frequent paths. & Basic performance measurements are already provided by the WFM/BPM system. \\
\hline
\end{tabular}
\vspace{-4mm}
\label{tab:processMiningTechniques}
\end{table*}

An important point is that also the traversal of gateways and internal or boundary events are included in the event log. So, not only the tasks are recorded, but the exact path of the model. This can simplify the
frequency or performance decoration of the process model: performing token-based replay or alignments to find the path that is followed is not necessary. A postprocessing activity is only necessary
when the execution of tasks needs to be analysed.

We will present the implementation of an connector in Section \ref{sec:setup}.
A property of the connector is that it is incremental: the first extraction extracts all the events from the beginning of the time,
while the following extractions extract only the events that are inserted since the previous extraction. This permits to keep the log updated, keeping
the workload low.

\subsubsection{Extracting Process Models}

Aside from event logs, we can also extract the BPMN models of the processes supported by the workflow engine.
In a Tomcat distribution of Camunda, each process supported by Camunda has its own folder in {\it PATH-TO-CAMUNDA-SERVER/webapps/}. As example, if a process has name {\it invoice}, its corresponding folder is
{\it PATH-TO-CAMUNDA-SERVER/webapps/invoice}. To extract the BPMN model associated to the invoice process, the content of the {\it PATH-TO-CAMUNDA-SERVER/webapps/invoice/WEB-INF/classes} folder should be taken.
As another possibility, we could refer to querying the REST API for the process diagram.

The extracted BPMN models can be imported in different process mining tools. In order to perform analyses such as decision mining and conformance checking (see Section \ref{sec:processMiningCamunda}),
the BPMN model should be converted to a Petri net model. This is difficult for many constructs
(for example, OR-joins and OR-splits, swimlanes, subprocesses) and thus can lead to problems for complex real-life processes.
An overview of the problematics of conversion from a BPMN model to a Petri net is found in \cite{dijkman2008semantics}.

\section{Process Mining on top of Camunda}
\label{sec:processMiningCamunda}

In the previous section, we have described an approach to extract event logs for the different processes supported by Camunda. This enables the application of several process mining
tools and techniques. In this section, we want to analyze which process mining techniques are most useful in the context of the Camunda processes.
The techniques are implemented and released as open-source software, including the one based on PM4Py presented in Section \ref{sec:setup}. Table \ref{tab:processMiningTechniques} provides an overview of
the approaches, along with their pros and cons.

\subsubsection{Process Discovery and Conformance Checking}
The two most popular process mining disciplines are process discovery and conformance checking. The scope of application of process discovery is pretty limited,
since the event data contained in the database is regulated by the process models inserted in Camunda.
A BPMN model is also a formal model that enables the application of conformance checking techniques.
For less regulated processes, the goal of conformance checking is to identify deviations in the process model, and the executions of the process are evaluated by their {\it fitness} according to the process model.
For WFM/BPM systems, we can expect to have perfect fitness for all the process executions. However, another application of conformance checking is the measurement of {\it precision}.
A model is precise when it does not allow for extra behavior, i.e., behavior that does not appear in the event data. Models can have low precision when they flexibly allow the execution sequence of activities. Hence, measuring precision
can provide a measure for the ``flexibility'' of the process model. A popular measure for precision is proposed in \cite{munoz2010fresh}.

\subsubsection{Decision Mining}
The application of a {\it decision mining} technique allows to enrich the model with {\it execution guards} that are extracted automatically from the event data. These are conditions that are required in order to execute
a path in the model.
Hence, decision mining helps to reduce the amount of behavior allowed in the process model by an adaptation of the model towards the guards that are discovered by the technique.
A mature approach for decision mining is proposed in \cite{de2013data}. On the other hand, BPMN models
are often already decorated with execution guards that are defined in the design phase.
Hence, decision mining could end up finding exactly the same guards without adding anything new.
Another problem is the discovery of trivial guards, or guards that overfit the data. A careful selection of the guards is necessary after performing the analysis.

\subsubsection{Concept Drift Analysis}
A {\it concept drift analysis} allows to identify the points in time where the execution of the process changes. Different types of concept drifts exist: sudden drifts (where the process becomes immediately significantly
different), gradual drifts and seasonal drifts. An approach for the detection of concept drifts is described in \cite{bose2011handling}. While the technique is interesting, many concept drift points are already known
in the context of WfMSs as Camunda:
for example, the underlying BPMN schema changes, or there are differences in the execution of a process between day and night.

\subsubsection{Prediction}
Given an incomplete process execution, it may be useful to estimate the remaining execution time based on historical executions. Several approaches have been proposed \cite{polato2018time,tax2017predictive}, however
in our experiments the quality of
predictions on top of real-life datasets is still not completely satisfying. Moreover, this paper only covers the extraction of complete (historical) process instances.

\subsubsection{Other Analyses}
In this category, we include {\it social network analysis} \cite{van2005discovering}, that with different metrics (such as the handover of work, the working together, the similar activities metric) calculates the collaboration
between the different organizational resources. {\it Model enhancement} with frequency and performance metrics is particularly important to identify the bottlenecks of the process (from a performance point of view)
and the most frequent paths.

\section{Set-Up of the Connector}
\label{sec:setup}

The connector presented in this section is completely open-source and is available at {\it https://github.com/Javert899/incremental-camunda-parquet-exporter}.
A completely working demo environment can be easily obtained by using {\it docker-compose} inside the folder of the project\footnote{The command {\it docker-compose up} starts the docker containers that are referred in the {\it docker-compose.yml} file}.
In the prepared environment, there are:
\begin{itemize}
\item A PostgreSQL relational database that is supporting the Camunda BPM engine and is exposed at port $5432$.
\item The Camunda BPM engine, that is running at port $8080$. The Camunda interface can be reached at \url{http://localhost:8080/camunda-welcome/index.html}. The installation contains some demonstrative models and event data that can be extracted by the connector and used for process mining analysis.
\item The connector, that is written in the Python language and is configured to reach the PostgreSQL database (for the extraction of the event data) and the Camunda BPM docker container (to extract the BPMN models).
\item An open-source process mining solution \cite{berti2019process,berti2019webservices}, with its own event logs database, that is reachable at port $80$, providing {\it admin}/{\it admin} as access credentials for the interface.
The services and the web interface are offering the logs for all the processes contained in Camunda. In the demo interface, a single process is offered to the user.
\end{itemize}

The processes contained in Camunda are offered, through the connector, in the web interface, graphically allowing for the following
operations:
\begin{itemize}
\item Process discovery of a directly-follows graph, and of a process tree or Petri net discovered using the inductive miner algorithm. While the model itself is already known,
the frequency and performance information are important to understand which parts of the process are more critical for key performance indicators and service level agreements.
\item Cases Exploration: understanding which cases have longer duration, and which are the events of such cases.
\item Social Network Analysis: shows the interaction between the organizational resources using the Camunda BPM engine through some classic metrics.
\end{itemize}
The deployment of the connector through {\it docker-compose} is integrated with the process mining tool, but the event log is available for usage
in other process mining platforms.
An example log that is extracted by the technique is available at \url{http://www.alessandroberti.it/invoice.xes}.

\section{Conclusion}
\label{sec:conclusion}

In this paper, we presented a tool to extract event logs and process models 
from the Camunda system and analyzed the applicability of process mining tools on such models and logs.
Thereby, process mining comes into reach of all organizations using Camunda with almost no effort.
The extractor we implemented was also integrated with the open-source process mining tool PM4Py, 
along with instructions on how to deploy a complete
environment that contains Camunda supported by the PostgreSQL database, the process mining tool, and the extractor.
The deployment shows how Camunda can be extended with process mining capabilities in a straightforward way.
While the PM4Py deployment is for demonstrative purposes, 
the extractor can be used on real-life deployments of Camunda
and combined with any process mining tool. An example use case for our tool is proposed at \url{http://www.alessandroberti.it/only_appendix.pdf}.

Next to  process discovery and conformance checking, our integration allows to discover the bottlenecks of the processes and to identify execution guards
that are not contained in the process model, but implicitly assumed by the resources performing the process. Thereby, the analysis can help to improve the quality of the BPMN model.
Other analysis, such as the detection of concept drifts, and the prediction of the remaining time, can also be useful.

As a result of this project, process mining techniques have been integrated in the {\it viadee Unternehmensberatung AG} software stack. The project located at \url{https://github.com/viadee/camunda-kafka-polling-client} proposes
an implementation of a polling client on top of the Apache Kafka event streaming platform to poll data from Camunda, and the project located at \url{https://github.com/viadee/bpmn.ai} proposes a data preparation pipeline for such process data.

\bibliographystyle{IEEEtran}
\bibliography{viadee}

\end{document}